\documentclass{aastex}
\usepackage[numberedappendix]{emulateapj5}
\usepackage{onecolfloat5}
\usepackage{fix2col}
\usepackage{amsmath,amssymb}
\usepackage{natbib}
\bibpunct{(}{)}{;}{a}{,}{,}

\input epsf

\def\plottwo#1#2{\centering \leavevmode
    \epsfxsize=1.\columnwidth \epsfbox{#1} \hfil
    \epsfxsize=1.\columnwidth \epsfbox{#2}}



\long\def\comment#1{}

\def\W2{{\cal W}}

\def\be{\begin{equation}}
\def\ee{\end{equation}}
\def\bea{\begin{eqnarray}}
\def\eea{\end{eqnarray}}

\def\cmm2{{\,\rm cm^{-2}}}
\def\cm2{{\,{\rm cm}^2}}
\def\cmm3{{\,{\rm cm}^{-3}}}
\def\gcmm3{{\,{\rm g\,cm^{-3}}}}

\def\fun#1#2{\lower3.6pt\vbox{\baselineskip0pt\lineskip.9pt
  \ialign{$\mathsurround=0pt#1\hfil##\hfil$\crcr#2\crcr\sim\crcr}}}

\newcommand{\half}{\ensuremath{\frac{1}{2}\,}}
\newcommand{\fsky}{\ensuremath{f_{\text{sky}}\,}}
\newcommand{\mth}{\ensuremath{m_{\text{th}}}}
\newcommand{\ngbar}{\ensuremath{\bar{n}_{g}}}
\newcommand{\Nbar}{\ensuremath{\bar{N}}}
\newcommand{\Nbarind}{\ensuremath{\bar{N}^{a}_{i}\,}}

\newcommand{\params}{\ensuremath{\Nbar^{a}_{i\alpha}}}

\def\phz{photo-{\it z}}
\def\trz{true-{\it z}}

\DeclareMathOperator{\erfc}{Erfc}

\newcommand{\figpath}{.}
\newcommand{\pspath}{.}
\newcommand{\bigfigsize}{1.0}

\lefthead{Schneider, Knox, Zhan \& Connolly} \righthead{Constraints on
  photo-z binned galaxy redshift distributions with two-point
  correlation functions}

\begin{document}
\bibliographystyle{apj} 
\twocolumn[
\submitted{To be submitted to ApJ} 
\title{Using Galaxy Two-point Correlation Functions to
Determine the Redshift Distributions of Galaxies Binned by
Photometric Redshift}
\author{Michael Schneider$^{1}$, Lloyd
Knox$^{1}$, Hu Zhan$^{1}$ and Andrew Connolly$^{2}$}
\affil{$^{1}$ Department of Physics, University of California, Davis,
CA 95616, USA \\ email: schneider@ucdavis.edu, lknox@ucdavis.edu,
zhan@physics.ucdavis.edu}
\affil{$^{2}$ Department of Physics and Astronomy, University of
Pittsburgh,
Pittsburgh, PA 15260, USA
\\ email: ajc@phyast.pitt.edu}
\begin{abstract}
We investigate how well the redshift distributions of galaxies sorted
into photometric redshift bins can be determined from the galaxy
angular two-point correlation functions.  We find that the uncertainty
in the reconstructed redshift distributions depends critically on the
number of parameters used in each redshift bin and the range of
angular scales used, but not on the number of photometric redshift
bins.  Using six parameters for each photometric redshift bin, and
restricting ourselves to angular scales over which the galaxy number
counts are normally distributed, we find that errors in the
reconstructed redshift distributions are large; i.e., they would be
the dominant source of uncertainty in cosmological parameters
estimated from otherwise ideal weak lensing or baryon acoustic
oscillation data.  However, either by reducing the number of free
parameters in each redshift bin, or by (unjustifiably)
applying our Gaussian analysis into the non-Gaussian regime, we find
that the correlation functions can be used to reconstruct the redshift
distributions with moderate precision; e.g., with mean redshifts
determined to $\sim 0.01$.  We also find that dividing the galaxies
into two spectral types, and thereby doubling the number of redshift
distribution parameters, can result in a reduction in the errors in
the combined redshift distributions.

\end{abstract}

\keywords{cosmology: theory -- cosmology: observation} ]

\section{Introduction}

There are many different techniques for determining the
distance--redshift relation and/or growth--redshift relation motivated
by the desire to understand the dark energy.  Those that rely on the
distances to a relatively small number of objects, such as the Type Ia
supernova method \citep[e.g.][]{riess98}, can use spectroscopic
redshift determinations and thus avoid redshift error as a significant
source of uncertainty.  However, when the distance (and/or growth)
constraints are derived from measurement of very large numbers of
objects spectroscopy can be a practical impossibility.  In such cases
one must rely on ``photometric redshifts''; i.e., redshifts estimated
from photometry in multiple broad bands
\citep[e.g.][]{loh86,connolly95,sawicki97}.

The relatively low cost per object of imaging surveys compared to
spectroscopic surveys is a great advantage and provides significant
motivation for pursuing the technique of estimating photometric
redshifts.  Imaging surveys can potentially constrain dark energy via
a variety of techniques including cluster counting
\citep[e.g.][]{haiman01}, cosmic shear \citep[e.g.][]{hu02d,huterer02}
and baryon acoustic oscillations
(BAO)~\citep[e.g.][]{seo03,blake03,padmanabhan06}.  It may even be possible for
imaging surveys to use Type Ia supernovae, without spectroscopic
follow-up, to constrain cosmology \citep{barris04}.

But abandoning spectroscopy has its disadavantages too.  In general,
there is some tolerance of redshift error, but less tolerance for
uncertainty about the probability distribution of those errors.  The
impact of redshift uncertainties on dark energy constraints has been
studied for supernovae \citep{huterer04}, cluster number counts
\citep{huterer04}, weak lensing
\citep{bernstein04,huterer06,ishak05,ma06} and baryon oscillations
\citep{zhan06,zhan06b}.

All of the studies cited in the above paragraph model the error
distribution as Gaussian. However, photometric redshift error
distributions, due to spectral-type/redshift degeneracies, often have
bimodal distributions, with one smaller peak separated from a larger
peak by $\Delta z$ of order unity
\citep[e.g.][]{benitez00,fernandez-soto01,fernandez-soto02}.  Thus a
fraction of galaxies have photometric redshifts that are
`catastrophically' wrong.  Here we study how well the coarse
properties of the true redshift distribution of galaxies in a given
photometric redshift bin can be reconstructed from galaxy
two-point correlation functions.

The idea is that catastrophic photometric redshift (``\phz{}'') errors
introduce additional correlations between galaxies in different
redshift bins.  In general, such errors will alter both the amplitude
and shape of the binned angular correlation functions.  Measurements
of the correlation functions over a range of angular scales would thus
provide valuable information to unravel the effects of large \phz{}
errors.

We emphasize that we are not attempting a forecast of the \phz{} errors
achievable given all possible information.  In particular, we neglect
information from spectroscopic calibration of the \phz{} error
distribution.  Spectroscopy, possibly combined with a ``super'' photometric
(12 or more bands) photo-z training set will play a critical role\footnote{
The current plan for \phz{} calibration is described in 
http://www.lsst.org/Science/photo-z-plan.pdf.}.  
In this sense our forecasts here are highly conservative.
Further, the
catastrophic errors are likely to be avoided by use of luminosity
function and surface brightness priors.  For recent results
on spectroscopic calibration of \phz{} measurements for weak lensing,
see \citet{ilbert06}.

The outline of this paper is as follows.  In \S~\ref{sc:method} we
describe our model for the \phz{} errors, the Fisher matrix we use to
constrain the parameters of this model, and our model for the galaxy
angular power spectra.  We present our results in
\S~\ref{sc:results}, including the details of our fiducial model and
its impact on the resulting Fisher matrix constraints.  We discuss
some implications of our results in \S~\ref{sc:discussion} and draw
conclusions on the feasibility of constraining \phz{} errors with
galaxy angular correlation functions.

\section{Method\label{sc:method}}
In this section, we first introduce our model for the catastrophic
\phz{} errors, and then describe the Fisher matrix
formalism~\citep{jungman96a,tegmark97c} that we use to forecast how
well the parameters of this error model can be constrained from
observations of the galaxy angular correlation function, binned in
redshift.  We restrict ourselves to forecasting here and leave for
later work the development of a practical algorithm for constraining
the \phz{} errors in a galaxy survey.

\subsection{Model for catastrophic photo-z errors}
To focus on the gross mislabeling of galaxy redshifts introduced by
catastrophic \phz{} errors, we bin the galaxy distribution in redshift
and model the errors as a linear mixing of the values of the galaxy
number density in each bin.  In terms of this model, our goal is then
to constrain the number of galaxies from each \trz{} bin that
contribute to the observed number in a given \phz{} bin.  

We assign the same numerical values for redshift intervals to \phz{}
bins and \trz{} bins. The parameters of our error model are then
defined as\footnote{We will use latin indices in the beginning of the
alphabet for the galaxy sub-populations, latin indices in the middle
of the alphabet for the observed redshift bins, and greek indices for
the true bins (the range of the two indices may be different in
general).}
:
\begin{description}
\item $\Nbar^{a}_{i\alpha}\equiv $ mean number of galaxies per
  steradian of spectral-type $a$ in \phz{} bin~$i$ that come from
  \trz{} bin~$\alpha$.
\end{description}
By considering only the \emph{mean} number of galaxies mixing between
\phz{} bins, we are ignoring possible angular fluctuations in the
mixing.  We expect this to be a good approximation on scales large
enough for the fractions of different galaxy types to be uniform, and
in the limit of homogeneous noise.  The separate index for galaxy
sub-populations is to allow for the possibility of different \phz{}
errors for different galaxy spectral types.  However, for most of our
results we ignore any information about galaxy types and consider just
the parameters $\Nbar_{i\alpha}\equiv\sum_{a} \Nbar^{a}_{i\alpha}$ of
the entire sample of galaxies.

Using these parameters, we construct the redshift distribution of
galaxies in \phz{} bin $i$,
\be\label{eq:photoz_dNdz}
  \frac{dN_{i}^{a}}{dz\,d\Omega}(z,\boldsymbol{\theta}) 
  = \sum_{\alpha} \Nbar^{a}_{i\alpha} 
  \left( \frac{1}{\Nbar^{a}_{\alpha}}\, 
  \frac{dN^{a}}{dz\,d\Omega}(z,\boldsymbol{\theta})\, 
  \psi_{\alpha}(z)\right)
\ee
where $dN^a/dzd\Omega$ is the number of galaxies of spectral-type $a$
in redshift interval $dz$ and angular interval $d\Omega$,
$\psi_{\alpha}(z)$ is a top-hat window function defining the \trz{} bin
$\alpha$, and
\be\label{eq:numdensity}
  \Nbar^{a}_{\alpha} = \frac{1}{\Omega}\int d\Omega\,
  \int_{0}^{\infty} dz\, 
  \frac{dN^{a}}{dz\,d\Omega}(z,\boldsymbol{\theta})
  \, \psi_{\alpha}(z)
\ee
is the mean number of galaxies (per steradian) in \trz{} bin $\alpha$.

Because we are binning in redshift, our model cannot tell us anything
about the shape of the redshift distribution within each bin (given by
the term in parentheses in eq.~\ref{eq:photoz_dNdz}).  We therefore
assume that this is known. However, the normalization of the redshift
distribution in each bin is determined by the parameters
$N^{a}_{i\alpha}$.

If we integrate eq.~\ref{eq:photoz_dNdz} over redshift, we get an
expression for the total number of galaxies (per steradian) in
each photo-z bin,
\be\label{eq:errormodel} 
  N^{a}_{i}(\boldsymbol{\theta}) = \sum_{\alpha}\frac{\Nbar^{a}_{i\alpha}} 
  {\Nbar_{\alpha}^{a}}\, N^{a}_{\alpha}(\boldsymbol{\theta})
\ee 
We take the set of $N^{a}_{i}(\boldsymbol{\theta})$ as our
data set that we use to constrain the mean parameters
$\Nbar^{a}_{i\alpha}$.

\subsection{Fisher matrix}
Rephrased in more abstract terms, our problem is to figure out how
well a set of parameters $\{a_p\}$ can be constrained from a data set
$\{N^a_i\}$, through the influence of $a_p$ on the statistical
properties of the data.

On large scales, the values of $N^a_i$ are Gaussian distributed and
their statistical properties are completely described by the mean,
$\Nbar^a_i$, and covariance, $w^{ab}_{ij}(\theta, a_p)$.  However, on small
scales, where nonlinear clustering becomes important, the galaxy
number density becomes significantly non-Gaussian and higher-order
correlations are required to completely describe the statistics of the
density field.  To avoid the complexities of non-Gaussianity, we
limit ourselves to a range in $\theta$ where the data is Gaussian to a
good approximation (see section~\ref{sc:fidmodel} for details). Note,
however, that there is more information in the data in smaller scales
than we are considering here, which  could improve the parameter
constraints beyond those shown below.


For Gaussian data, the Fisher matrix is given by~\citep{tegmark97c},
\be\label{eq:Fisher} 
  F_{pp'} = \frac{\partial \mathbf{\Nbar}^{T}}{\partial
    a_{p}} \mathbf{w}^{-1} \frac{\partial\mathbf{\Nbar}}{\partial
    a_{p'}} +\half\text{tr}\left( \frac{\partial\mathbf{w}}{\partial
    a_{p}} \mathbf{w}^{-1} \frac{\partial\mathbf{w}}{\partial a_{p'}}
  \mathbf{w}^{-1}\right) 
\ee 
where $\mathbf{\Nbar}$ is the mean of the data and $\mathbf{w}$ is the
covariance matrix of the data, defined by, 
\be 
  \mathbf{N} =
  \mathbf{\Nbar} + \delta\mathbf{N}, 
\ee 
with $\mathbf{N}\equiv \{N_{i}^{a}\}$, and 
\be 
  \mathbf{w} = \left<\delta\mathbf{N}\,
  \delta\mathbf{N}^{T} \right>
\ee 
so that $\mathbf{w} = \left\{ w^{ab}_{ij}(\theta,a_{p})\right\}$.  In
the quadratic approximation to the likelihood, the inverse Fisher
matrix is then equal to the covariance of the parameters $a_{p}$.

In this paper, we consider two parameter sets for $\{a_p\}$.  First,
we calculate the Fisher matrix for the parameters $\{\params\}$.  We
then add the linear galaxy bias with respect to the dark matter in
each redshift bin ($\{b^{a}_{\alpha}\}$) and the amplitude of the
power spectrum at $z=0$ (which is degenerate with the galaxy bias at
$z=0$) as additional parameters.

The expression for the Fisher matrix in eq.~\ref{eq:Fisher} simplifies
considerably when we Fourier transform over the variable $\theta$.  In
this case, the covariance matrix becomes block-diagonal (with one
block for each value of the conjugate variable $\ell$) due to
isotropy.  The second term in eq.~\ref{eq:Fisher} then becomes,
\begin{multline}\label{eq:Fisher2}
  F^{(2)}_{pp'} = \half\fsky \sum_{ \substack{ \ell \\(ai)(ai)'\\
  (bj)(bj')}} (2\ell+1)\\ 
  \times\left[ \frac{\partial\,C_{(ai)(ai)'}}{\partial a_{p}} 
  \left(\mathbf{C}^{-1}\right)_{(ai)'(bj)'} 
  \frac{\partial\,C_{(bj)'(bj)}}{\partial
  a_{p'}} \left(\mathbf{C}^{-1}\right)_{(bj)(ai)} \right],
\end{multline}
where $\fsky$ is the fractional angular area of the sky covered by the
survey, $C_{(ai)(bj)}(\ell)$ is the Fourier transform of
$w^{ab}_{ij}(\theta)$ and we have written out the matrix
multiplications explicitly\footnote{We treat each pair of indices
$(ai)$ as a single index on a two-dimensional matrix for the power
spectrum between redshift bins for each galaxy sub-population.}.

When we Fourier transform the first term in eq.~\ref{eq:Fisher}, only
the monopole ($\ell=0$) contributes to the mean so that,
\be\label{eq:Fisher1} 
F^{(1)}_{pp'} = \sum_{(ai),(bj)}
\frac{\partial\Nbar^{a}_{i}}{\partial a_{p}}
\left(\mathbf{C}^{-1}(\ell=0)\right)_{(ai)(bj)}
\frac{\partial\Nbar^{b}_{j}}{\partial a_{p'}} 
\ee 
This term tells us what information the mean of the data contributes
to the parameter constraints.  Because $\mathbf{\Nbar}$ does not
depend on the galaxy bias, eq.~\ref{eq:Fisher1} is nonzero only for
the parameters $\{\params\}$.

While formally the cosmological
monopole of the power spectrum cannot be determined (because of
unknown contributions from super-horizon-sized modes), for a survey
that covers only a fraction of the sky, this term will be aliased and
will contain contributions from the power spectrum at small (but
nonzero) $\ell$.  We therefore define an ``effective'' monopole
variance,
\be\label{eq:monopole}
  C_{(ai)(bj)}(\ell=0) \equiv \delta^{K}_{(ai)(bj)}
    \left(\sigma^{0}_{(ai)}\right)^{2}
\ee
From the predicted amplitude of the angular power
spectrum at small $\ell$ we set $\sigma^{0}_{(ai)} = 10^{-3}\times
\Nbar_{i}^{a}$. We consider this a conservative upper bound.

To avoid confusion in the strict interpretation of this term, we point
out that it is identical in form to adding extra Fisher information
from an independent measurement of $\mathbf{\Nbar}$.

Note that, in our notation, \citet{ma06} and \citet{zhan06b} have
omitted the term in eq.~\ref{eq:Fisher1} and instead fixed
$\Nbar_{\alpha} \left(= \sum_{i} \Nbar_{i\alpha}\right)$.  By letting
$\Nbar_{\alpha}$ vary freely, we are assuming no prior knowledge of
the true redshift distribution.  And by including the term in
eq.~\ref{eq:Fisher1}, we are taking into account the fact that the
mean number of galaxies in each \phz{} bin can be determined from the
data.  \citet{huterer06} and \citet{zhan06} use a method more similar
to ours in this respect; i.e., they fix the total number of galaxies
in each photometric redshift bin and allow $\Nbar_{\alpha}$ to vary.
\citet{huterer06} find there is very little difference between the two
approaches for the case of weak lensing.  This may not be the
case though for galaxy power spectra.  

\subsection{Galaxy angular power spectrum with \phz{} errors}

\begin{figure*}[htpb]
\plottwo{\pspath/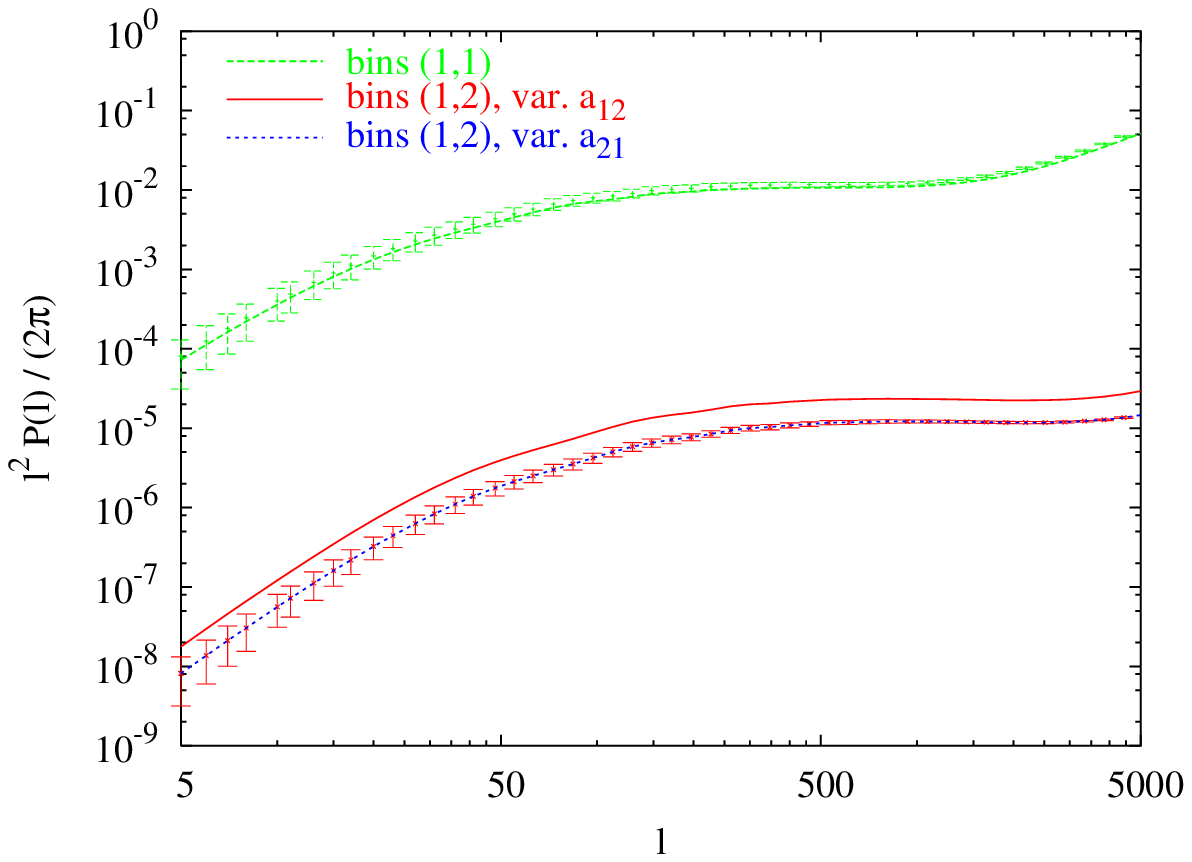}{\pspath/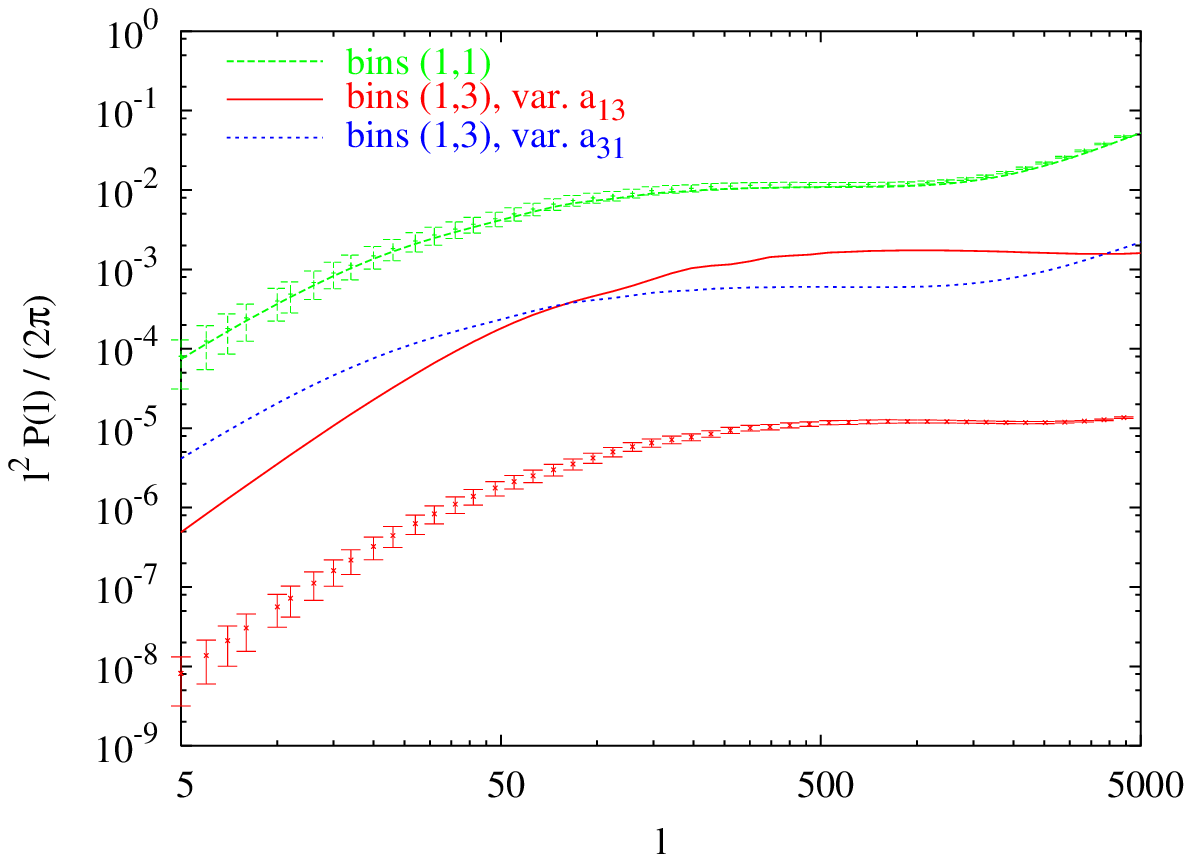}
\caption{\label{fg:ps} Angular galaxy auto and cross power spectra for
  \phz{} bin~1 in our fiducial model.  The points with error bars are
  the fiducial power spectra, while the lines are 1-$\sigma$
  variations of the parameters $a_{12}\equiv \Nbar_{12}$ or
  $a_{21}\equiv \Nbar_{21}$ (left) and $a_{13}\equiv
  \Nbar_{13}$ or $a_{31}\equiv \Nbar_{31}$ (right).  For
  clarity, we have shown only the cross power spectrum that varies the
  most when the parameters are varied, with the auto power spectra
  (fiducial points and variation) for reference.  The 1-$\sigma$
  variations of the parameters are from the Fisher matrix (see
  fig.~\ref{fg:LSSTbias} below).}
\end{figure*}

Using eq.~\ref{eq:errormodel}, the observed angular power spectrum is
related to the ``true'' angular power spectrum (i.e. without \phz{}
errors) by,
\be\label{eq:psobs}
  C_{(ai)(bj)}(\ell) = 
  \sum_{\alpha\alpha'}\,
  \Nbar^{a}_{i\alpha} \Nbar^{b}_{j\alpha'}\,
  P^{ab}_{\alpha\alpha'}(\ell),
\ee
where bars denote quantities averaged over a redshift bin and
$P_{\alpha\alpha'}^{ab}(\ell)$ is the power spectrum of the normalized
density fluctuations $\delta N^{a}_{\alpha}/\Nbar^{a}_{\alpha}$ (while
$C_{(ai)(bj)}(\ell)$ is the power spectrum of $\delta\Nbar^{a}_{i}$).  We
also add shot noise to the model for the observed power spectrum,
\[
  C^{\text{noise}}_{(ai)(bj)} = \delta^{K}_{(ai)(bj)}
  \frac{\Nbarind}{ A_{\text{survey}}},
\]
where $A_{\text{survey}}$ is the angular area of the survey in square
arcminutes and $\delta^{K}$ is the Kronecker delta function. We plot
fiducial power spectra given by eq.~\ref{eq:psobs} in fig.~\ref{fg:ps}
along with the variation of the power spectra when one of the
parameters $\params$ is changed by 1-$\sigma$.

We use the halo model to obtain analytic expressions for the linear
galaxy bias and nonlinear 3-D galaxy power spectrum as described in the
appendix.  We then use the Limber approximation~\citep{limber53} to
project this into the binned angular galaxy power spectrum,
\be\label{eq:angPS}
P_{\alpha\beta}^{ab}(\ell) =
\delta^{K}_{\alpha\beta}\frac{2\pi^{2}}{\ell^{3}} \int_{0}^{\infty} dz
\frac{d\chi}{dz} \left(\Delta^{2}\right)^{ab}_{\alpha\beta}
\left(\frac{\ell}{\chi},z\right) \chi\, W_{\alpha}^{a}(\chi)\,
W_{\beta}^{b}(\chi) 
\ee 
where $\chi(z)$ is the comoving angular
diameter distance as a function of redshift, $\Delta^{2}(k)$ is the
3-D variance of galaxy number density fluctuations per logarithmic
interval in $k$, and
\[
  W_{\alpha}^{a}(\chi) \propto \frac{d\bar{N}^{a}}{d\chi} \psi_{\alpha}(\chi)
\]
is the probability distribution for finding a galaxy of type $a$ in
$z$-bin $\alpha$ at a comoving distance $\chi$ in the survey, \emph{in
the absence of \phz{} errors}, with $\psi_{\alpha}(\chi)$ a top-hat
window function for the $z$-bin $\alpha$, as defined in
eq.~\ref{eq:numdensity}, and normalization $\int
W_{\alpha}^{a}(\chi)\,d\chi=1$.  To simplify the computation, we
ignore the redshift evolution of $\Delta^{2}(k,z)$ in
eq.~\ref{eq:angPS}; evaluating it instead at the mean redshift of the
bin.

Note the delta-function in eq.~\ref{eq:angPS} so that our model
neglects any intrinsic cross-correlation between redshift bins.  For
$\ell=50$ (near the peak of the angular power spectrum) and redshift
bins with width of 0.5, we have checked that the cross-correlations
between bins are less than 1\% of the auto-correlations.  However,
with our fiducial model, the catastrophic \phz{} errors induce
correlations at the level of $\sim5$\% of the auto-correlation (see
fig.~\ref{fg:ps}, left panel).  We therefore expect that neglecting
intrinsic cross-correlations should not alter our Fisher matrix errors
by more than $\sim 1\%$, which is completely unimportant for the
conclusions we draw here.

An important effect that we have neglected here is magnification of
galaxies by weak lensing from intervening large-scale
structure~\citep{turner80,turner84}.  
Lensing can induce correlations between different \phz{} bins beyond
those that would be found in an unlensed model.  The amplitude of the
weak-lensing induced correlations depends on the luminosity function
of the galaxies in the survey, but could be of similar magnitude to
the \phz{} error induced correlations~\citep{villumsen95}.  We expect
this effect can be calculated with sufficient accuracy that residual
uncertainties will not lead to significant confusion with photo-z
error induced correlations.

\subsection{Galaxy bias parameterization}

If we restrict ourselves to linear theory, then the galaxy power
spectrum factors into a product of the dark matter power spectrum
times a constant (scale-independent but redshift dependent) bias:
\be\label{eq:biasPS} 
  P^{\text{gal,lin}}_{(a\alpha)(b\beta)}(\ell) =
  b_{\alpha}^{a}\, b_{\beta}^{b}\,
  P^{\text{DM,lin}}_{(a\alpha)(b\beta)}(\ell).  
\ee 
At small scales, the
bias $b_{\alpha}^{a}$ becomes scale-dependent and this factorization
no longer holds.  Therefore, when we include the parameters
$\{b_{\alpha}^{a}\}$ in our Fisher matrix analysis, we are
approximating the galaxy power spectrum with the linear power
spectrum.  Below, we evaluate the effects of this assumption.

\subsection{Galaxy sub-populations}

We consider two ways of utilizing the multiband photometric data in a
galaxy survey.  On the one hand, we imagine assigning \phz's to each
galaxy and binning them according to their \phz's while ignoring any
other features of the galaxies.  We can then use the number counts in
each bin along with our model for the power spectrum to constrain the
true redshift distribution of all the galaxies in each bin.  On the
other hand, we can sort the galaxies according to their spectral (or
morphological) type and repeat the analysis (jointly) for each
sub-population.  From this larger data set, we can then infer
constraints on the parameters $\Nbar_{i\alpha}\equiv \sum_{a} \params$
for the full galaxy sample by adding the variances of the $\params$
(including cross-terms)\footnote{This is equivalent to making a formal
change of variables in the inverse Fisher matrix from $\params$ to
$\Nbar_{i\alpha}$}.  We expect the $\params$ to be different for each
sub-population not only because of different redshift distributions
for each galaxy type, but also because of different \phz{} error
distributions.

While the shot noise increases by dividing the galaxy sample into
sub-populations, it may be possible to improve the constraints on the
parameters of the total galaxy sample for several reasons.  First, we
gain knowledge of the mean number density of each sub-group, which
contributes to the term in eq.~\ref{eq:Fisher1}.  Second, we gain
extra information from the cross-correlation between the sub-groups,
which is even more helpful if the redshift distribution of one
of the types can be well-constrained on its own.  In fact, the
exposure times and bands for the survey could even be optimized for
the best constrained galaxy sub-population.  In the limit of exact
redshifts of a sub-sample of galaxies,~\citet{newman06}
has shown that cross-correlating with the \phz{} sample can place
significant constraints on the \phz{} errors.  Third, a given galaxy
sub-group may be more biased than the total galaxy sample and could
therefore have equivalent or greater S/N than the total sample even
though the noise increases for the sub-samples.  

\subsection{Mean redshift in each \phz{} bin\label{sc:zbar}}
As discussed in the introduction, we would like to know if the
constraints on the \phz{} error parameters shown in
fig.~\ref{fg:LSSTbias} will be sufficient to enable interesting
constraints on dark energy parameters from weak lensing and baryon
acoustic oscillation surveys.  To properly address this question, we
should forecast joint constraints on a suite of cosmological and \phz{}
parameters with both galaxy and shear data.  However, this analysis
has already been done for the case of Gaussian \phz{}
errors~\citep{zhan06b} and we choose not to repeat that effort here.
Instead, we make contact with previous work by reducing the
constraints on the $\params$ to constraints on the mean redshift in
each \phz{} bin, defined as,
\be
  \bar{z}^{a}_{i} \equiv \frac{\int dz\, z\, d\Nbar^{a}_{i}/dz}
  {\int dz\, d\Nbar^{a}_{i}/dz}.
\ee
The errors on $\bar{z}_{i}^a$ are extracted from the inverse Fisher
matrix,
\be\label{eq:sigmazbar}
  \left(F^{-1}_{\bar{z}}\right)_{(ai)(bj)} = \sum_{(ck\beta),(ck\beta)'} 
  \frac{\partial\bar{z}^a_i}{\partial \Nbar^c_{k\beta}}\, 
  \frac{\partial\bar{z}^b_j}{\partial \Nbar^{c'}_{k'\beta'}}\,   
  \left(F^{-1}_{\Nbar}\right)_{(ck\beta)(ck\beta)'}
\ee
where $F_{\Nbar}$ is the Fisher matrix in eq.~\ref{eq:Fisher}.

It has been shown in~\citet{huterer06,ma06,zhan06,zhan06b} that
useful dark energy constraints require the error on the mean redshift
in each \phz{} bin to be constrained to $\sim0.003$ near $z \simeq 1$,
and increasing rapidly towards higher redshift.  We therefore use
0.003 as a benchmark for assessing our results.

\section{Results\label{sc:results}}

In this section, we first describe a simple fiducial model for the
galaxy survey and \phz{} errors, and then show the forecasted
constraints from the Fisher matrix in eq.~\ref{eq:Fisher2}.  We then
study how our results depend on the fiducial model in order to draw
conclusions independent of the many assumptions in our model.

\subsection{Fiducial model\label{sc:fidmodel}}

We choose our fiducial model to mimic the LSST\footnote{www.lsst.org}.
Throughout, we consider a survey covering 20,000~sq.~deg. and bin the
galaxies in \phz{} over the range 0 to 3.  Our fiducial cosmological
parameters are $\Omega_m=0.24$, $\Omega_{b}h^{2} = 0.022$,
$\Omega_{\Lambda}=0.76$, $h = 0.72$, and $\sigma_{8}=0.74$.

To ensure that our assumption of Gaussianity in
eq.~\ref{eq:Fisher} is reasonable, we limit the $\ell$ range in
eq.~\ref{eq:psobs} by adding large noise to each element of
$P_{(ai)(bj)}(\ell)$ with $\ell > \ell_{\text{max}}(z)$, where
$\ell_{\text{max}}(z) \equiv \chi(z)\,k_{\text{max}}(z)$
and\footnote{``DM'' denotes the dark matter power spectrum}
$\Delta^{2}_{\text{DM}}(k_{\text{max}},z) = 0.4$.
We also set a lower bound on $\ell$ to justify our use of the Limber
approximation for the angular power spectrum in eq.~\ref{eq:angPS}.
The Limber approximation relies on the observation that, when
projecting the 3-D power spectrum into 2 dimensions, the dominant
contribution is from those Fourier modes that do not oscillate
significantly along the line of sight.  We can quantify this statement
by considering only modes with line-of-sight component of the
wavevector $k_3 < 2\pi/\Delta\chi(z)$, where $\Delta\chi(z)$ is the
comoving width of the $z$-bin under consideration.  The Limber
approximation also requires $\ell\gg k_{3}\chi(z)$.  Putting these
together, we set $\ell_{\text{min}}(z)\equiv 4\pi \chi(z)/\Delta\chi$ (with
an arbitrary factor of 2 inserted just to be conservative).
\begin{table}
\caption{Multipole range for the galaxy power spectrum as a function of
  redshift\label{tb:ellmax}}
\centering
\begin{tabular}{ccc}
\tableline
\tableline
$z$ range & $\ell_{\text{min}}(z)$ & $\ell_{\text{max}}(z)$\\
\tableline
0.0 -- 0.5 & 7  & 114 \\
0.5 -- 1.0 & 23 & 458 \\
1.0 -- 1.5 & 45 & 1018 \\
1.5 -- 2.0 & 71 & 1875 \\
2.0 -- 2.5 & 103 & 3195 \\
2.5 -- 3.0 & 140 & 5186 \\
\tableline
\end{tabular}
\end{table}
In table~\ref{tb:ellmax}, we show the values of $\ell_{\text{max}}$,
as a function of redshift in our fiducial model for 6 $z$-bins over
the range $0\le z \le 3$.  We evaluate both $\ell_{\text{min}}$ and
$\ell_{\text{max}}$ at the centers of the \phz{} bins.


We use the redshift distribution from~\citet{song03} (which is based
on Subaru observations with limiting magnitude in $R$ of 26),
\be\label{eq:dNdz} 
\frac{d\Nbar}{dz\,d\Omega}(z) = \Nbar_{\text{tot}}\,\exp\left[-\left(
\frac{z}{1.2} \right)^{1.2} \right] \times
  \begin{cases}
    z^{1.3} & z<1 \\
    z^{1.1} & z>1
  \end{cases}
\ee 
with the normalization, $\Nbar_{\text{tot}}$ set by $\int
dz\,d\Nbar/dz\,d\Omega=65$ per~sq.~arcmin.

When considering galaxy sub-populations, we consider two spectral
types that we label ``red'' and ``blue,'' roughly depending on the
absence or presence of active star formation.
In the absence of a well-motivated model,
we generated several redshift
distributions for the red and blue sub-populations in an ad-hoc
fashion, and compared the results between them.  We require only that
the redshift distributions sum to give eq.~\ref{eq:dNdz} and that the
blue distribution dominate at large redshifts ($z \gtrsim 1$).

\begin{figure}[htpb]
\centerline{
\scalebox{0.7}{\includegraphics{\figpath/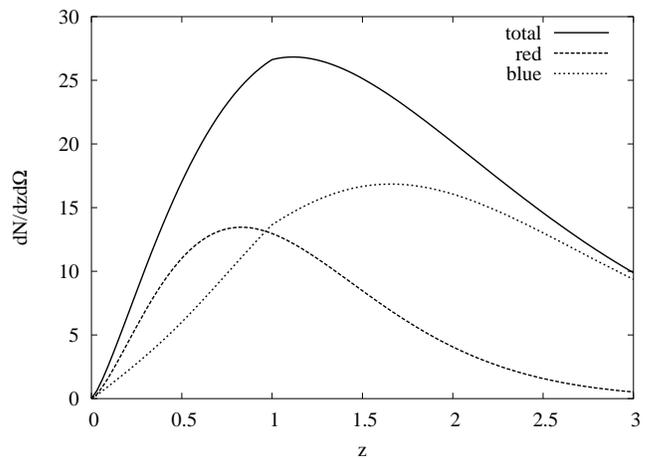}}
}
\caption{\label{fg:dNdz}Fiducial redshift distributions for the
  ``red'' and ``blue'' galaxy sub-populations (eq.~\ref{eq:dNdz_red})
  and the total galaxy sample (``total,'' eq.~\ref{eq:dNdz}).}
\end{figure}

For the red and blue galaxy sub-populations, we set
\be\label{eq:dNdz_red}
  \frac{d\Nbar^\text{red}}{dz\,d\Omega} = R\,\Nbar_{\text{tot}}\,z^{1.4}\,
  \exp\left[-r_{1}z^{r_{2}}\right] 
\ee
and 
\[
  \frac{d\Nbar^{\text{blue}}}{dz\,d\Omega} = 
  \frac{d\Nbar}{dz\,d\Omega}-\frac{d\Nbar^{\text{red}}}{dz\,d\Omega}
\]
with $R=0.8$,
$r_1=1.3$, $r_2=1.4$.  These fiducial redshift distributions are
shown in fig.~\ref{fg:dNdz}.

Our fiducial model for $\params$ is based on the estimated \phz{}'s
for a simulated random sample of 100,000 galaxies over redshifts from
0 to 4 with colors assigned by filtering spectra from a sample of 10
redshift-evolved spectral energy distributions (SEDs).  The simulation
assumed photometric data was available in 6 filters ($ugrizy$),
modelled after the LSST, with the data limited in the $i$-band at
$i<25$ and an S/N of 10-15 at the depth of the survey. The depth of
the simulation is what one would achieve after about 400 visits per
filter.  This is just a fiducial approach as the errors can be
optimized by weighting the exposure times in each band separately.
The \phz{} of each galaxy was estimated by matching the galaxy colors
with a SED template library.  No priors on the luminosity function or
surface brightness were used, which can significantly improve the
\phz{} estimates in some cases.  In this regard, our fiducial model is
therefore a worst-case scenario.

To model the errors for the galaxy
sub-populations, we divided the templates for the simulated galaxy
SEDs into 2 groups based on the presence or absence of strong emission lines.

\begin{figure*}[!htpb]
\plottwo{\figpath/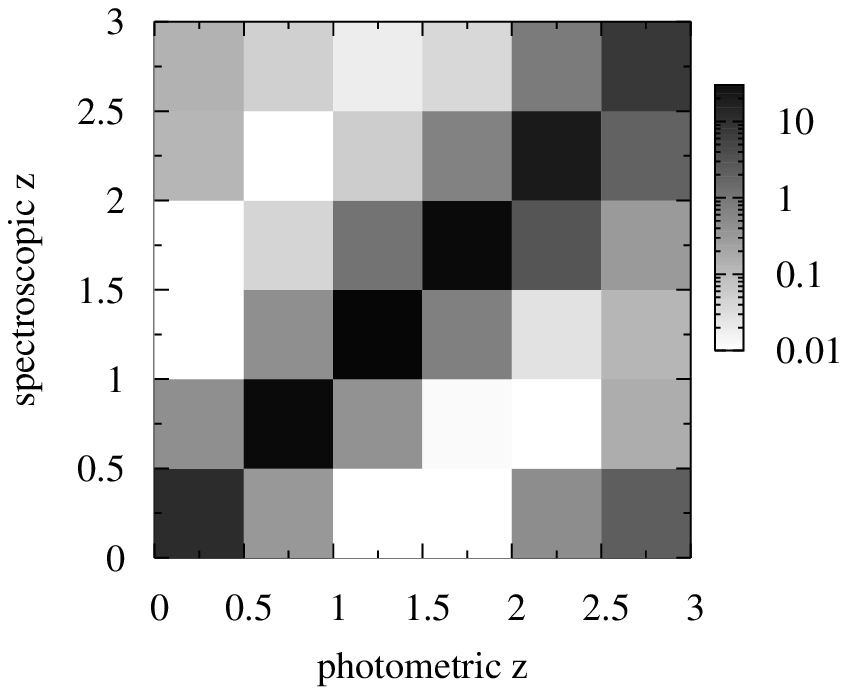}{\figpath/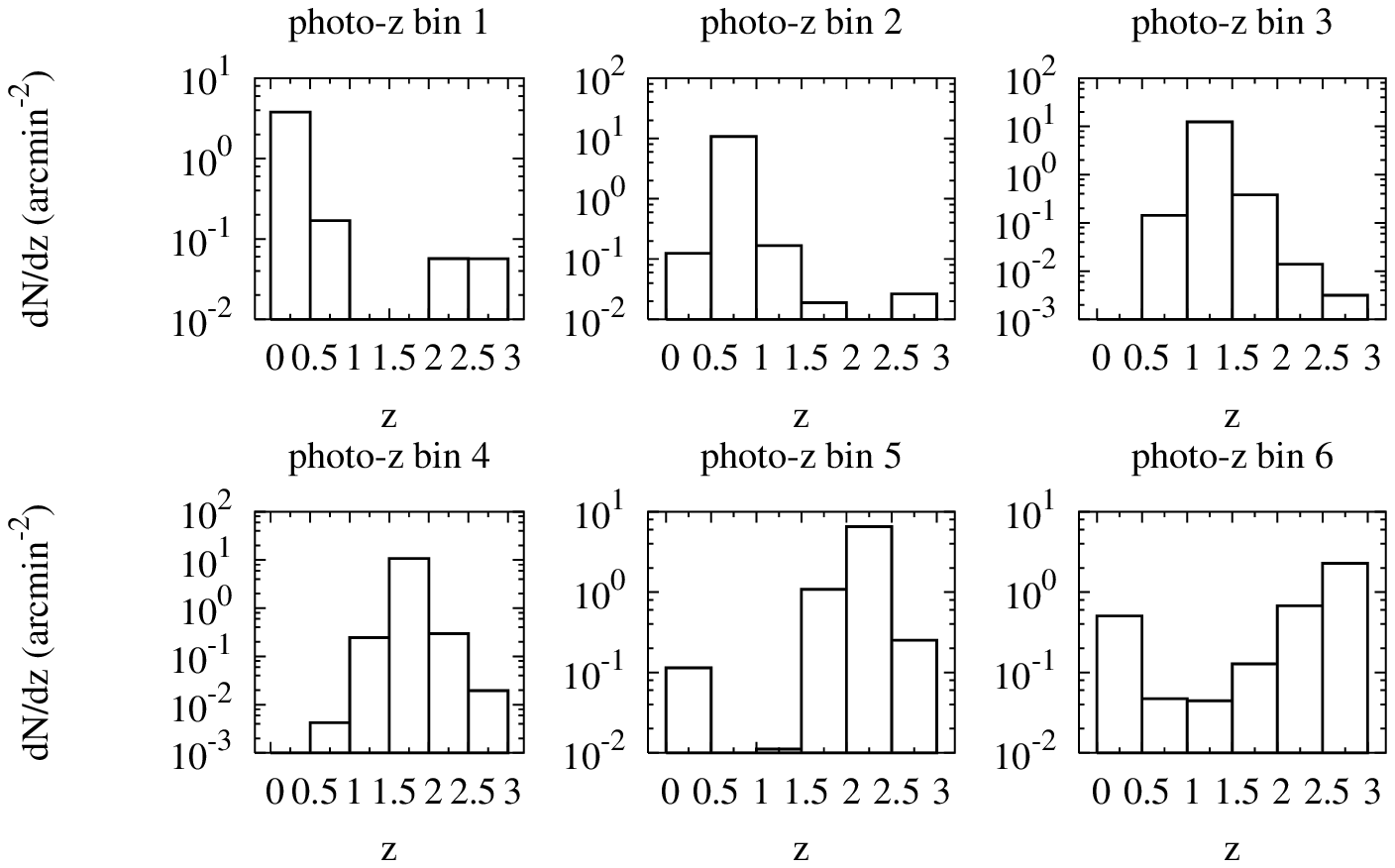}
\caption[]{\label{fg:fid_dNdz} The fiducial redshift distribution with
catastrophic \phz{} errors: $\params$ (for the ``LSST'' model -- see
text).  The two plots show the same parameters in different
perspectives.  On the left, is the number density, $\params$, as a
function of redshift (indexed by $i$~=~photometric $z$ and
$\alpha$~=~spectroscopic $z$).  On the right, each window shows a
different \phz{} bin index, $i$, while each bar in a given window
shows a different spectroscopic index, $\alpha$, for the given $i$.
For example, the bar between $z=0.5$ and $z=1$ in the window for
``\phz{} bin 1'' is the number density of galaxies with spectroscopic
redshifts in the range $0.5<z<1$ that have been given photometric
redshifts in the range $0<z<0.5$.}
\end{figure*}

The fiducial parameters, $\params$, are constructed by first creating
the matrix, $E^{a}_{i\alpha} \equiv \params / \Nbar^{a}_{\alpha}$,
by binning the \phz{} vs. $z$ plane 
and normalizing so that $\sum_{i}E^{a}_{i\alpha}=1$ (for each
$a$ and $\alpha$).  This normalization conserves the total number of
galaxies in the survey.  We then use the fiducial redshift
distribution in eq.~\ref{eq:dNdz} to create $\Nbar^{a}_{\alpha}$
according to eq.~\ref{eq:numdensity} and multiply with
$E^{a}_{i\alpha}$ to get the parameters $\params$.
An example of our fiducial model for the 
$\params$ is plotted in fig.~\ref{fg:fid_dNdz}.

\subsection{Parameter constraints\label{sc:constraints}}
\begin{figure*}[htpb]
\centerline{\scalebox{\bigfigsize}
{\includegraphics{\figpath/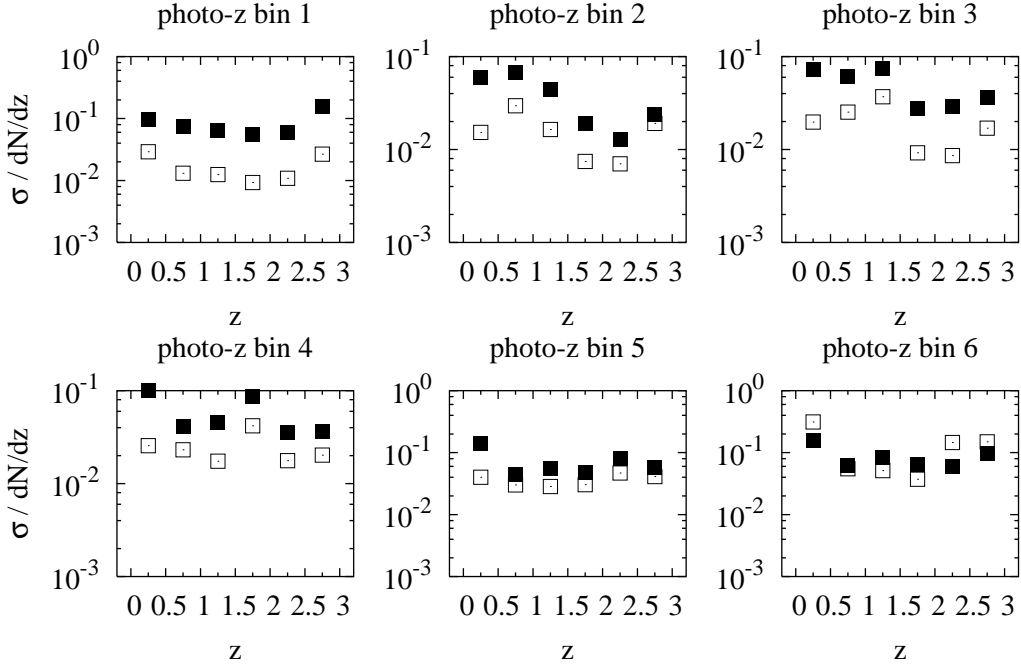}}}
\caption{\label{fg:LSSTbias} Forecasted constraints for the parameters
$\Nbar_{i\alpha}$ divided by the fiducial $\Nbar_i$ in each \phz{}
bin, {\it i.e.} the error on the fraction of outliers within each
\phz{} bin.  The filled squares are the fractional constraints when
no galaxy sub-populations are considered, while the open squares
are the constraints when the galaxy sample is divided into ``red'' and
``blue'' spectral types.  For the open squares, the parameters for
the ``red'' and ``blue'' sub-populations ($\params$) were constrained
first, and then these constraints were combined to produce the
constraints on $\Nbar_{i\alpha}$ shown here (by summing over the index
$a$ in the inverse Fisher matrix components).  A 10\% prior on the
galaxy bias and a 100\% prior on the $\Nbar^{a}_{i\alpha}$ were
imposed.  The fiducial model assumed ``LSST'' \phz{} errors (see text)
and sky coverage of 20,000~sq.~deg.}
\end{figure*}

Our main results are in fig.~\ref{fg:LSSTbias}, which shows the Fisher
constraints on the parameters $\Nbar_{i\alpha}$ assuming a 10\% prior
on the galaxy bias and 100\% prior on the $\params$.  The model for
the open squares includes the ``red'' and ``blue'' galaxy
sub-populations, while the model for the filled squares ignores this
information.  

In the column~2 of table~\ref{tb:zbar}, we show the constraints
on the mean redshift in each \phz{} bin (eq.~\ref{eq:sigmazbar})
implied by the constraints on the $\params$ without galaxy
sub-populations (filled squares in fig.~\ref{fg:LSSTbias}).  These are
two orders of magnitude larger than the ``benchmark'' value (described
in section~\ref{sc:zbar}) needed for constraining dark energy
parameters.  In the following subsections, we discuss three ways that
the constraints on $\bar{z}_i$ could possibly be improved.
\begin{table}
\caption{Constraints on the mean redshift of each \phz{} bin\label{tb:zbar}}
\centering
\begin{tabular}{ccccc}
\tableline
\tableline
$z$ range & $\sigma_{\text{LSST}}$ & $\sigma_{\text{sub}}$ &
$\sigma_{\text{Gauss}}$ & 
$\sigma_{\ell_{\text{max}}=4000}$\\
\tableline
0.0 -- 0.5  & 0.29 & 0.059 & 0.0050 & 0.0030 \\
0.5 -- 1.0  & 0.068 & 0.041 & 0.010 & 0.0050 \\
1.0 -- 1.5  & 0.12 & 0.036 & 0.0087 &  0.0096 \\
1.5 -- 2.0  & 0.16 & 0.049 & 0.0091 & 0.016 \\
2.0 -- 2.5  & 0.23 & 0.085 & 0.0079 & 0.022 \\
2.5 -- 3.0  & 0.28 & 0.61 & 0.0047 & 0.025 \\
\tableline
\end{tabular}
\end{table}

\subsubsection{Adding galaxy sub-populations}

\begin{figure*}
\centerline{
\scalebox{\bigfigsize}
{\includegraphics{\figpath/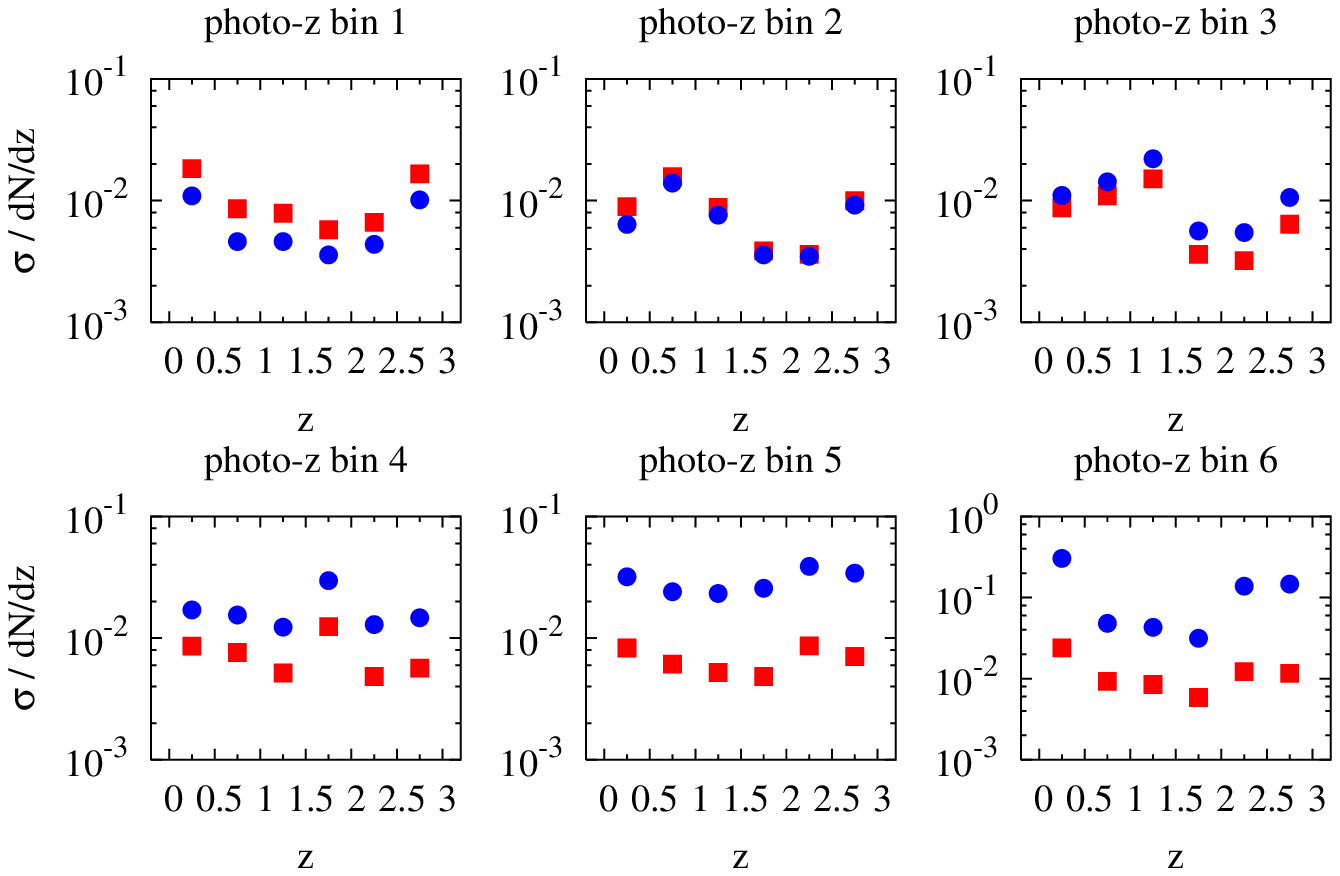}}
}
\caption{\label{fg:redblue} Forecasted fractional constraints on
  $\params$ for the ``red'' (squares) and ``blue'' (circles) galaxy
  sub-populations. These constraints take into account the
  cross-correlation between the red and blue samples.  We used a 10\%
  prior on the linear galaxy bias and a 100\% prior on the
  $\params$. }
\end{figure*}

For a wide range of fiducial models, we find that dividing our galaxy
sample into ``red'' and ``blue'' sub-populations improves the
forecasted constraints on the redshift distribution in each \phz{}
bin, as shown by the open squares in fig.~\ref{fg:LSSTbias} and
column~3 of table~\ref{tb:zbar}.  We also show forecasted constraints
on the redshift distributions of the sub-populations in
fig.~\ref{fg:redblue}.  Comparing the errors on the mean redshift in
each \phz{} bin in table~\ref{tb:zbar}, we see that there is a
significant improvement over the constraints obtained without using
information about galaxy sub-populations, but is still much larger
than the ``benchmark'' for dark energy surveys given in
section~\ref{sc:zbar}

Because our fiducial models for the biases and redshift distributions
of the sub-populations (in the absence of \phz{} errors) are rather
ad-hoc, we have varied the parameters in eqs.~\ref{eq:dNdz_red} and
\ref{eq:HODsat} over a range of physically reasonable values and
find no change to the qualitative nature of our results.  This is
discussed more in section~\ref{sc:bias_sensitivity}.

\subsubsection{Sensitivity to parameterization}

We show in the left panel of fig.~\ref{fg:gauss_lzmax} and column~3 of
table~\ref{tb:zbar} the forecasted parameter constraints with a
fiducial \phz{} error model that only allows mixing between adjacent
\phz{} bins and with the number of \phz{} error parameters reduced to
only those that can take nonzero values in the fiducial model.  The
fiducial errors assume a 5\% contribution from each of the adjacent
bins to a given \phz{} bin.  This crudely mimics a Gaussian model for
the \phz{} errors.  We have also tightened the prior on the galaxy
bias from 10\% to 1\%.  While the constraints on the $\params$ in the
left panel of fig.~\ref{fg:gauss_lzmax} are moderately improved from
the default model in fig.~\ref{fg:LSSTbias}, the constraints on
$\bar{z}_{i}$ in table~\ref{tb:zbar} improve by nearly two orders of
magnitude.  This shows that reducing the number of free parameters in
each \phz{} bin indeed has a large impact on the ability to constrain
the redshift distribution in each bin.

\subsubsection{Sensitivity to range of angular scales}

The forecasted constraints are very sensitive to the maximum $\ell$
used in the galaxy power spectrum (but are rather insensitive to the
minimum $\ell$ cutoff).  In particular, for the lowest \phz{} bin, the
maximum cutoff at $\ell=114$ (from table~\ref{tb:ellmax}) removes some
of the baryon features in the power spectrum that can help in
diagnosing \phz{} errors~\citep{zhan06b}.

\begin{figure*}[htpb]
\plottwo{\figpath/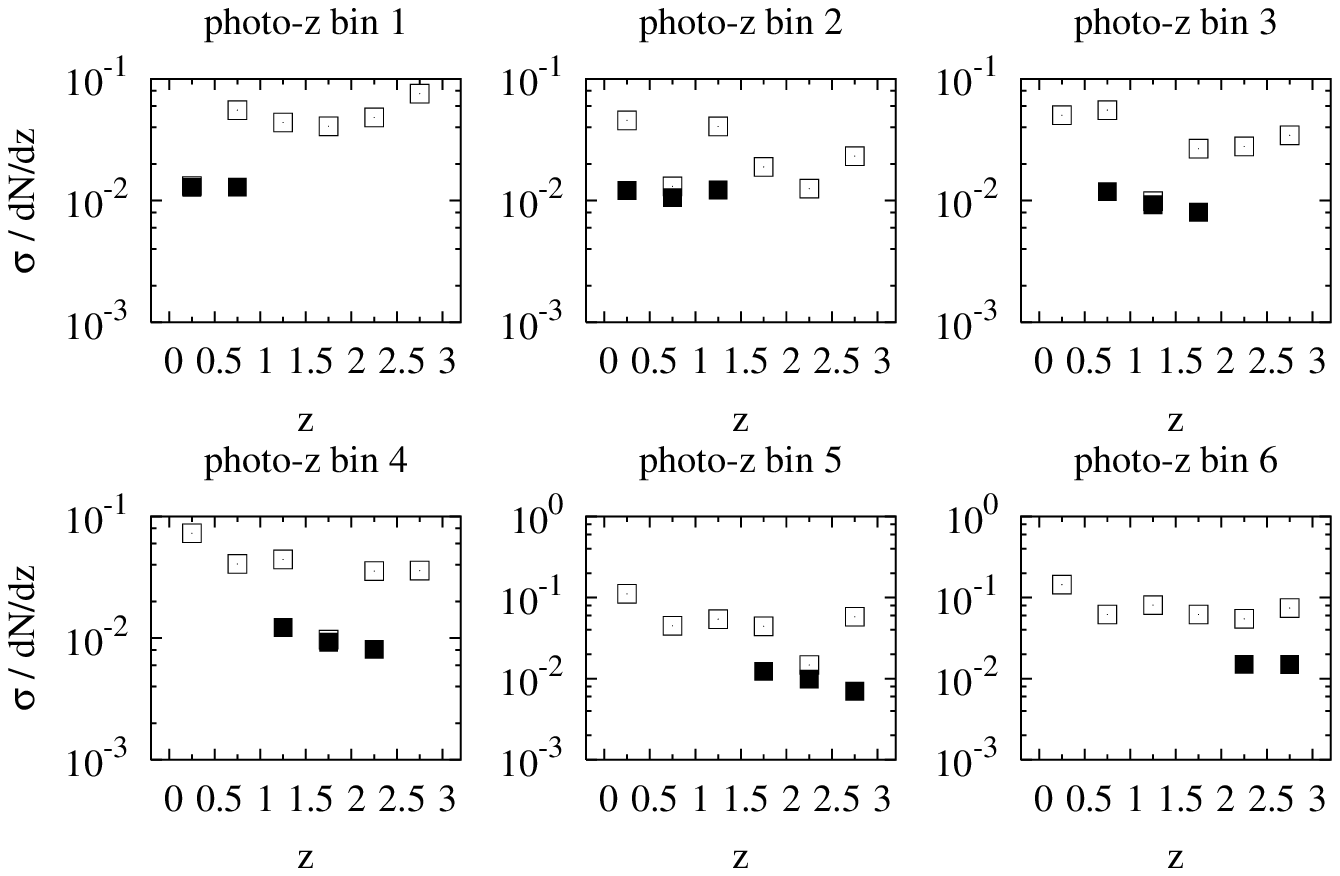}
{\figpath/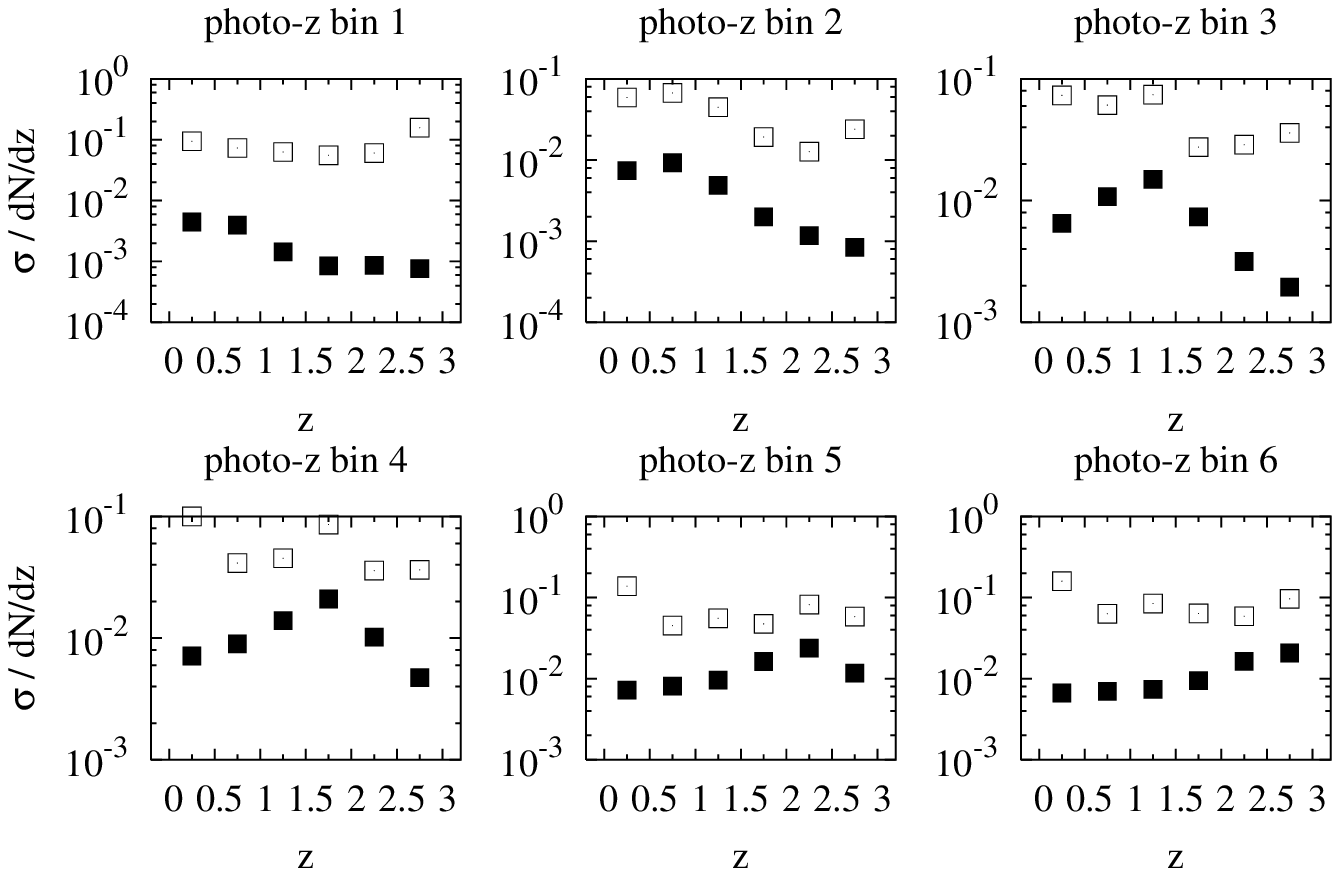}
\caption{\label{fg:gauss_lzmax} Demonstration of two factors that
  significantly affect the errors on the mean redshift in each \phz{}
  bin (see table~\ref{tb:zbar}).  For the filled squares on the left,
  we have chosen a fiducial \phz{} error model that only introduces
  mixing between adjacent bins and have fixed the parameters that are
  known to be zero in this model.  This roughly mimics a Gaussian
  \phz{} error distribution.  For the filled squares on the right, we
  have replaced the $z$-dependent $\ell$ cutoff from
  table~\ref{tb:ellmax} with a maximum $\ell=4000$ for all the \phz{}
  bins.  For reference, the open squares in both plots show the
  constraints with the default fiducial model without galaxy
  sub-populations.  The left panel assumes a
  1\% prior on the linear galaxy bias, while right panel assumes a
  10\% prior.}
\end{figure*}

To demonstrate this sensitivity, we show the forecasted constraints
when $\ell_{\text{max}}(z) = 4000$ for all $z$ in the right panel of
fig.~\ref{fg:gauss_lzmax} and column~5 of table~\ref{tb:zbar}.  The
constraints on the mean redshift in each \phz{} bin are two orders of
magnitude smaller than those with $\ell_{\text{max}}$ from
table~\ref{tb:ellmax} (labelled $\sigma_{\text{LSST}}$ in
table~\ref{tb:zbar}).

Recall that the maximum $\ell$ cutoff is imposed to validate our
assumption of Gaussian data (in e.g. eq.~\ref{eq:Fisher}).  Therefore,
the constraints presented in this section should be interpreted only up
to non-Gaussian corrections, which could be quite large.  Our results
are an indication that there is much to be gained by developing the
appropriate tools for analyzing the non-Gaussian case.

\subsection{Fiducial model dependence\label{sc:modeldep}}
To test the robustness of our conclusions, we recomputed the Fisher
matrix in eq.~\ref{eq:Fisher2} while varying the number of redshift
bins, the galaxy bias and halo occupation distribution in the
nonlinear power spectrum, and the fiducial model for the \phz{}
errors.

\subsubsection{Redshift distribution}
For comparison, we use a second fiducial model for $E^{a}_{i\alpha}$
with 10\% of the galaxies in each \phz{} bin uniformly distributed
over the remaining \phz{} bins.  This model, though not physically
motivated, gives us a reference for determining the sensitivity of the
Fisher constraints to the fiducial error model.

\begin{figure}[!htpb]
\centerline{
\scalebox{0.7}{\includegraphics{\figpath/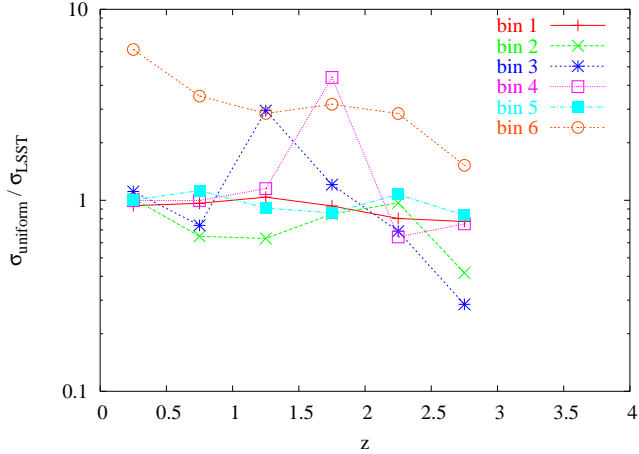}}
}
\caption[]{\label{fg:fiderror_compare} Ratios of forecasted errors for
the parameters $\Nbar_{i\alpha}$ using 2 different fiducial models for
the \phz{} errors.  We first scale the forecasted errors by the
fiducial $\Nbar_i$ in each \phz{} bin (to obtain fractional errors)
and then divide the errors obtained using a uniform distribution of
10\% scatter from each \phz{} bin by the forecasted errors obtained
using a \phz{} error model based on the LSST (see text).  Each line
shows a different \phz{} bin (corresponding to the index $i$ in the
parameters).}
\end{figure}

We show ratios of the Fisher matrix errors obtained with these two
fiducial \phz{} error models in fig.~\ref{fg:fiderror_compare}.  For
most of the parameters, the different fiducial models lead to
differences in the forecasted errors of a factor of $\lesssim 5$.  The
LSST fiducial model for the \phz{} errors is actually well-motivated
by our \phz{} estimation simulation so any uncertainties in the
fiducial model will be much smaller than the changes introduced by
this artificial ``uniform'' error model.  We therefore conclude that
uncertainties in the fiducial \phz{} errors will affect our forecasted
constraints by factors of $\lesssim$ a few.  

\subsubsection{Galaxy bias and nonlinear power spectrum 
\label{sc:bias_sensitivity}}
\begin{figure}[!htpb]
\centerline{
\scalebox{0.7}
{\includegraphics{\figpath/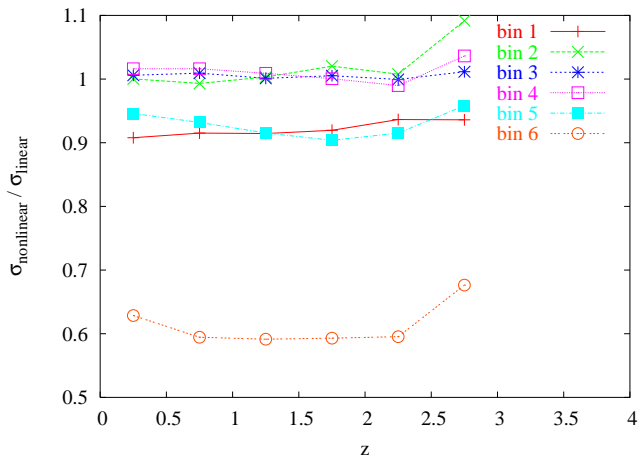}}
}
\caption{\label{fg:lincompare} Ratios of forecasted constraints for
the parameters $\Nbar_{i\alpha}$ using the nonlinear power spectrum
divided by the constraints using the linear theory power spectrum.
The fiducial redshift distribution and \phz{} errors are the same in
each case.  Each line shows a different \phz{} bin (corresponding to
the index $i$ in the parameters).}
\end{figure}

The results we have shown so far model the galaxy power spectrum using
only the linear theory prediction.  To make sure that this
approximation will not affect the qualitative nature of our results,
we compare in fig.~\ref{fg:lincompare} the forecasted errors for the
\phz{} error distribution obtained using the linear theory power
spectrum to those obtained with the nonlinear model (see the
appendix).  We see that our use of the linear power spectrum is a good
approximation, which is to be expected given our truncation in $\ell$
described at the beginning of section~\ref{sc:fidmodel}.

The fiducial linear galaxy bias and nonlinear galaxy power spectrum
depend on a model for the way that galaxies populate dark matter
halos.  The details of our fiducial model are explained in the
appendix, section~\ref{sc:HOD}, but this model has only been very
loosely constrained by observations~\citep{cooray06,abazajian05d}.  
Therefore to build confidence in our use of a necessarily ad hoc
fiducial model, we compared the parameter constraints from the Fisher
matrix when we vary the ad hoc parameters.  We find that when 
the ``satellite'' galaxy normalization and slope
(defined in eq.~\ref{eq:HODsat}) are varied by 50\% and 25\%
respectively, the changes in the \phz{} error parameter constraints
are less than 10\% and 0.01\% for all the redshift bins.  This amount
of variation will not affect our conclusions.


\subsubsection{Number of \phz{} bins}
We have compared the fractional constraints on the \phz{} error
parameters when the number of \phz{} bins is varied from 2 to 10 and
do not find a significant variation.  As the number of \phz{} bins is
increased, there is more information about the \phz{} errors from the
additional cross-correlations between \phz{} bins, but the number of
parameters to constrain also increases.  So, the fractional
constraints we show here for six bins should be representative of the
constraints that would be obtained for any moderate number of bins.
Note, however, that having the same \emph{fractional} constraints for
a larger number of parameters means we have more information about the
\phz{} error distribution with \phz{} bins.  Of course, for a
sufficiently large number of \phz{} bins, the shot noise will begin to
dominate.



\section{Discussion and Conclusions\label{sc:discussion}}

We have shown that the ability to constrain general
(i.e. non-Gaussian) \phz{} error distributions with galaxy two-point
correlation functions depends on the parameterization of the \phz{}
errors, the range of angular scales probed by the correlation
function, and prior knowledge of the galaxy bias.

Binning the galaxy sample in \phz{}, we have presented constraints on
the binned redshift distribution and mean redshift in each \phz{} bin.
Parameterizing the redshift distribution by binned values is
insensitive to small scatter from \phz{} errors, but otherwise assumes
no {\it a priori} knowledge of the \phz{} error distribution.  We find
that reducing the number of parameters in each \phz{} bin can be very
helpful, which could be achieved with improved knowledge of the \phz{}
errors from, e.g., spectroscopically calibrated samples or luminosity
function priors.

We have limited our use of the galaxy correlation function to angular
scales where the galaxy number density is Gaussian distributed.  At
low redshifts, this severely limits the amount of data
available to constrain the \phz{} error parameters.  We hypothesize
that including information from correlations on non-Gaussian scales
could significantly improve the constraints and demonstrate that the
constraints on the mean redshift in each \phz{} bin do improve by two
orders of magnitude with a naive extrapolation of our Gaussian
calculation to non-Gaussian scales.

If it is possible to separate the galaxies by spectral type, the
constraints on the \phz{} errors may improve further by including
information from the cross-correlation of the galaxy sub-samples.  We
have demonstrated this in figs.~\ref{fg:LSSTbias} and \ref{fg:redblue}
by separating our fiducial galaxy sample into ``red'' and ``blue''
spectral types.  We expect this procedure to be particularly helpful
if there exists a well-populated spectral class of galaxies whose
\phz{}'s can be estimated unusually well.

Our forecasts are limited to parameters of the \phz{} error
distribution and linear galaxy bias so we cannot make any rigorous
conclusions about what kind of dark energy constraints can be achieved
in weak lensing and baryon acoustic oscillation surveys with the level
of \phz{} errors forecasted here.  However, we make qualitative
comparisons with dark energy forecasts in the
literature~\citep{huterer06,ma06,zhan06,zhan06b} using our
constraints on the mean redshift in each \phz{} bin given in
table~\ref{tb:zbar}.  In the Gaussian regime, the constraints we
forecast of $\sim0.01$ are factors of a few larger
than those desired for upcoming dark energy surveys.  However, adding 
non-Gaussian scales in the correlation function may provide the
required constraints.   The galaxy correlation properties are quite
likely to provide at least a powerful consistency test for the redshift
distributions as determined via spectroscopic and/or ``super'' (12 or
more band) calibration subsamples.

\acknowledgments We thank M.~Auger, G.~Bernstein, D.~Huterer, D.~Koo,
J.~Newman, and J.~A.~Tyson for useful conversations .  This work was
supported in part by NSF grant 0307961.

\appendix
\section{Halo model}

The halo model~\citep[for a review see][]{cooray02} provides an
analytic approximation to the nonlinear three-dimensional galaxy power
spectrum using the assumptions that all the dark matter is contained
in gravitationally bound ``halos'' of varying mass and that the number
of galaxies populating a given dark matter halo is determined solely
by the halo mass and redshift.  The model for the number of galaxies
in a dark matter halo is often referred to as the ``halo occupation
distribution'' (HOD).

\subsection{Fiducial HOD models\label{sc:HOD}}
Following~\citet{hu04b}, we divide the mean number of galaxies in a
dark matter halo of mass $m$ into contributions from a galaxy at the
halo's center ($\Nbar_c$) and satellite galaxies ($\Nbar_s$).  The
mean number of central galaxies is essentially a unit step function
parameterized by a minimum threshold mass, $\mth(z)$, for a halo to
host a galaxy.  To allow for scatter in the relation between galaxy
luminosity and halo mass, the simple step function is modified to,
\be
  \label{eq:HODcen} \Nbar_c^a(m,z) = \half f^{a}(m,z)\,\erfc\left(
  \frac{\log(\mth(z)/m)}{\sqrt{2}\,\sigma} \right), 
\ee 
where
$f^{a}(m,z)$ is the fraction of central galaxies of spectral type $a$.
We use eq.~7 in~\citet{cooray06} as our fiducial model for
$f^{a}(m,z)$ and set $\sigma=0.1$.

The mean number of satellite galaxies is modelled as a power law,
\be\label{eq:HODsat}
  \Nbar_{s}(m,z) = \left( \frac{m}{A\,\mth(z)}\right)^{b},
\ee
with the two free parameters $A\sim30$, $b\sim1$~\citep{hu04b}.

The threshold mass, $\mth$, is determined by requiring the HOD model
to reproduce the fiducial redshift distribution as follows:
\begin{align}\label{eq:mthdef}
  \frac{d\Nbar}{dz\,d\Omega}(z) &= \chi^{2}(z)\frac{d\chi}{dz}\, 
  \int dm\, n(m,z)
  \left( \Nbar_{c}(m,z) + \Nbar_{s}(m,z)\right)\notag\\
  &\equiv \chi^{2}(z)\frac{d\chi}{dz}\,\ngbar(z),
\end{align}
where $\chi(z)$ is the comoving distance as a function of redshift and
$n(m,z)$ is the halo mass function (we use the Sheth-Tormen model for
$n(m,z)$~\citep{sheth99}).  

\subsection{Galaxy power spectrum}
The power spectrum of galaxies in the halo model is the sum of two
terms:
\[
  P_g(k,z_1,z_2) = P_{1h}(k,z_1,z_2) + P_{2h}(k,z_1,z_2),
\]
where,
\begin{multline*}
  P_{1h}(k,z_1,z_2) \equiv \frac{\delta_{z_1,z_2}}{\ngbar^{2}(z_1)}
  \int dm\, n(m,z)\\
  \times \left(\Nbar_s^2(m,z_1)\,u_g^2(k|z_1,m) +
  2\Nbar_c(m,z_1)\Nbar_s(m,z_1)u_g(k|z_1,m) \right)
\end{multline*}
is the contribution to the power from a single
halo\footnote{$\delta_{z_1,z_2}$ is the Kronecker delta function}, and 
\be\label{eq:P2h}
  P_{2h}(k,z_1,z_2) \equiv P^{\text{lin}}(k,z_1,z_2)\, I_2(k,z_1)\, I_2(k,z_2)
\ee
with,
\begin{multline*}
  I_2(k,z)\equiv \frac{1}{\ngbar(z)} \int dm\, n(m,z)\, b_{h}(m,z) \\
  \times\left(\Nbar_c(m,z) +\Nbar_s(m,z)u_g(k|z,m)\right)
\end{multline*}
is the contribution from 2 different halos.  Here, $u_g(k|z,m)$ is the
Fourier transform of the galaxy number density profile (assumed to
follow the NFW profile~\citep{navarro96}), $b_h(m,z)$ is the halo bias
(specified in the Sheth-Tormen model along with the mass function) and
$\ngbar(z)$ is the mean comoving number density of galaxies defined in
eq.~\ref{eq:mthdef}. 

\subsection{Linear power spectrum}
The linear power spectrum in eq.~\ref{eq:P2h} is the variance
per logarithmic interval in $k$, in linear perturbation theory
\[
  P^{\text{lin}}(k,z_1,z_2) \equiv \delta_{H}^{2}(0)\,g(z_1)\, g(z_2)
  \left(\frac{ck}{H_0}\right)^{n+3} T^{2}(k)
\]
where $\delta_H(0)$ is the amplitude at $z=0$, $g(z)$ is the linear
growth function, $T(k)$ is the transfer function, and $H_0$ is the
Hubble constant.  We take a fiducial value of $n=1$.

\subsection{Linear galaxy bias}
The galaxy bias in the halo model is given by,
\be\label{eq:galbias}
  b^{a}\left(z_{\alpha}\right) = \int dm\, n(m,z_{\alpha})\,
  b_{h}(m,z_{\alpha})\, \frac{\left(\Nbar^{a}_{c} +
    \Nbar^{a}_{s}\right)(m,z_{\alpha})}{\ngbar^{a}(z_{\alpha})},
\ee
where the superscript ($a$) labeling different galaxy sub-populations
denotes different values of the parameters $A$ and $b$ in
eq.~\ref{eq:HODsat} and $\mth$.

\bibliography{cmb4}

\end{document}